**Seeking instructional specificity: an example from analogical instruction**

Eric Kuo & Carl E. Wieman
Department of Physics and Graduate School of Education
Stanford University
Stanford, CA

**Abstract**

Broad instructional methods like "interactive engagement" have been shown to be effective, but such general characterization provides little guidance on the details of how to structure the instructional materials. In this study, we seek instructional specificity by comparing two ways of using an analogy to learn a target physical principle: (i) applying the analogy to the target physical domain on a *Case-by-Case* basis and (ii) using the analogy to create a *General Rule* in the target physical domain. In the discussion sections of a large, introductory physics course (N = 231), students who sought a *General Rule* were better able to discover and apply a correct physics principle than students who analyzed the examples *Case-by-Case*. The difference persisted at a reduced level after subsequent direct instruction. We argue that students who performed *Case-by-Case* analyses are more likely to focus on idiosyncratic problem-specific features rather than the deep structural features. This study provides an example of investigating how the specific structure of instructional materials can be consequential for what is learned.

**I: Introduction**

Physics Education Research has produced instructional strategies that have been categorized as "active learning" or "interactive engagement" [1]. One main contribution of PER has been to show that these instructional strategies can lead to gains for a variety of valued measures as compared to traditional instruction [2].

At the same time, not all interactive engagement is equally successful at achieving these goals [3,4]. The broadly defined instructional principle of interactive engagement leaves many free parameters up to the discretion of the instructor. Prior studies have shown there is significant variation in how instructors interactively engage students, even when nominally using the same instructional approach [5,6]. We have been investigating a different source of instructional variance: the structure of the instructional materials.

"Interactive engagement" as a teaching principle does not help one decide how to best structure the materials with which students are engaging to maximize learning. In this work we compare two different activity structures and show that there is a clear difference, a difference that can be understood in terms of studies of learning from cognitive psychology. We use this difference to illustrate the untapped benefits of seeking this level of specificity in instructional recommendations.

**II: Structuring Analogical Instruction in Physics**

This work considers the specific example of teaching the relationship between electric field and electric potential with an analogy to topographical contour maps.

Physics instruction regularly draws on analogies to teach new and unfamiliar concepts. For example, flowing water through pipes leads to circuit concepts, waves on a string connect to electromagnetic waves, and topographical contour maps relate to equipotential diagrams for electric charges.

The use of analogies to understand a new system in terms of a known one is a key part of professional scientific practice and discovery [7–9]. It is therefore not surprising that the use of analogies to teach introductory physics concepts can help students learn important features of more abstract physical phenomena [10–13]. For example, in an effort to teach students about the normal force that rigid objects can apply, Brown and Clement [12] use the idea of springs exerting a force back on your hand when compressed to show how rigid objects like tables can exert a normal force on objects.

Existing traditional and PER-inspired instructional materials often incorporate analogies into their design [14,15]. For example, *Physics by Inquiry* prompts students to build analogies between different physical quantities (such as density and heat capacity) or between different situations (such as hot air rising and a piece of wood floating to the surface of the water). These types of activities have been successful at increasing the conceptual learning that can be achieved in physics instruction. But while the choice of which analogy used for teaching a physical system has been shown to be consequential for what students learn [10,11], an unanswered question explored in this work is how the structure of analogical instruction in physics is consequential for what students learn.

There are at least two possible instructional approaches to teaching physics principles in our target physical domain from an analogical domain. One approach is to provide students with the task of mapping an analogy into several situations in the target physical domain, with the expectation they will learn from this repeated practice. We refer to this as a *Case-by-Case* instructional approach.

A second approach is to give students the explicit task of using the reasoning from the analogical domain to develop a general rule or explanation that will apply to all situations in the target physics domain. We refer to this as a *General Rule* instructional approach.

This study investigates how the difference between these two instructional approaches, *Case-by-Case* (CC) and *General Rule* (GR), is consequential for student learning outcomes. We found that students instructed through the *General Rule* approach are more successful than those instructed through the *Case-by-Case* approach at discovering and applying the relation between electric field and electric potential.

*A: Case-by-Case: Learning by Applying to Specific Examples*

Instructional materials designed for "interactive engagement" commonly have students actively work through a series of specific questions to support the development of student reasoning, inquiry, and conceptual understanding. The design of these questions typically relies on research on common student difficulties [16–19]. Here, in the *Case-by-Case* approach, students were presented with an analogy and were asked to use it to answer specific questions in the target physical domain. Feedback came from a computer simulation where students were able to check their answers. The sequence of questions was designed to provide students initially with simpler problems, and later, more complex ones. This was designed to give students some basic practice in using the analogy before exploring more difficult examples.

On each of the individual questions, students were provided with the explicit hint to recall and apply the analogy. Previous research has shown that spontaneous mapping from a single analogical context to a novel one is rare and that an explicit hint to recall the analogy is more successful in prompting analogical thinking [20–22]. Explicitly cued analogical thinking can lead to later spontaneous analogical thinking in a new context when these explicit cues are not present [23].

*B. General Rule: Learning through Generalization*

Students in the *General Rule* approach were given the task of using the analogy to create a general rule that could apply across all situations in the target physical domain. After some guidance on how to connect the contour map analogy to electric potential lines, they were given freedom to create these general rules with little explicit guidance on what features to consider. These students also had access to the computer simulation for developing and getting feedback on their rules. This instruction was designed to support the learning of domain-relevant, general relations that could be directly used to answer questions in this target domain.

By seeking a general rule that can apply to all cases, rather than answers to questions for particular cases, students may seek the deep structure of the analogy [20,24]. In other contexts, the task orientation of inventing general explanations has been shown to have a benefit for identifying the deep structural features of a phenomenon [25–28]. The task here of developing a general rule for all cases of a physical principle may similarly help students avoid distraction from case-specific details.

Another possible benefit of the *General Rule* instructional approach is that the immediate task is framed as applicable to future situations. By expansively framing the task as creating a general rule that can be used to solve future problems, students may expect that their rule should be applied in future settings [29].

**III: Learning Hypotheses**

While we expect both groups to learn, cognitive arguments suggest there will be differences:

1) *GR better supports discovery and application the correct principle than CC*: There are two reasons to anticipate the comparative benefits of the *General Rule* approach over the *Case-by-Case* analysis: better generalization, and the analogy is more salient. Consideration of individual cases in isolation can lead to development of idiosyncratic, situation-specific rules which do not hold true more generally [30]. Directing students to analyze individual situations in the *Case-by-Case* instruction may have the unintended effect of suppressing the need for a general explanation that holds across all the cases. Expansively framed as developing a general explanation that can be used to solve all future problems, the *General Rule* instruction may push students to articulate what is important to take from the analogy, promoting identification of a single, common explanation and better recognition of the important underlying structure in the target physical domain.

Another possible benefit of the general rule instruction is that analogical mapping decreases significantly when not explicitly prompted [20,22,31]. In the *Case-by-Case* activity, students are repeatedly cued to map an analogical situation into a novel one. When these cues later disappear, so may students' reasoning with those analogies. Students directed to develop a general rule *in the context of the target domain* might shorten the distance between the analogical reasoning and the target domain, leading to increased salience and use of the analogy. This increased salience and use may also be supported by an expansive framing that the previously generated rule is relevant for later problems.

2) *Competing predictions of the effect of subsequent direct instruction*
What effect will subsequent direct instruction after these different activities have on this conditional difference? In typical physics courses, even activities emphasizing discussion of student ideas will be followed by subsequent direct instruction on the relevant physics principles. We came up with two plausible, but competing, hypotheses of the effect of subsequent direct instruction.

First, a common instructional intuition would suggest that giving all students direct instruction on the correct principle cancels out any initial post-activity differences in recognition and application of the correct rule. This is consistent with a perspective that views direct instruction as superior to more weakly-guided exploration that relies on potentially weak problem-solving strategies [32].

In contrast, previous research has also shown that even unsuccessful student effort to create a general explanation can prepare students to learn from future direct instruction [25–27,33]. Schwartz and Bransford [33] showed that students who analyzed a relevant data set before direct instruction of psychological principles were better able to make predictions with those principles for a hypothetical research study. They argue that the preparatory data analysis helped students differentiate key features of the phenomenon, such that later learning of relevant conceptual frameworks of this phenomenon is enhanced. If similar arguments apply here, this would favor GR over CC.

The inclusion of subsequent direct instruction in our study design tests these two competing hypotheses: whether subsequent direct instruction will enhance or nullify a difference between the two instructional activities.

**IV: Study Design**

*A: Research Context and Participants*

This research study was conducted at an elite private university. Participants were enrolled in a large lecture, calculus-based, introductory physics course covering electricity and magnetism. Total enrollment in this course was about 500 students. The course meeting times consisted of 50-minute lectures, meeting three times a week, and a 50-minute discussion section that met once a week. There were 32 discussion sections, and enrollment for each discussion section was capped at 18. This course is primarily taken by engineering majors, and most students were 1$^{st}$ or 2$^{nd}$ year undergraduate students.

The main intervention occurred in the discussion sections, each led by a teaching assistant (TA).  Of the 16 TAs, who each taught 2 discussion sections, 8 taught with the GR materials and 8 taught with the CC materials.  The TAs were assigned to the activities such that TA gender distribution and teaching experience were about equivalent in the two instructional conditions.  Before instruction, the research team led a TA training meeting.  The two groups of TAs were split into separate rooms, where they discussed the overall purpose of their instructional activity, became familiar with their activity, and discussed pedagogical suggestions and potential student pitfalls.  The research team emphasized to both groups that the TAs should facilitate the discussion sections by leading students to think through the analogy and not to simply provide students with answers to the worksheet.  The TAs were aware that two different versions of the activities were being used and studied but were blind to the researchers' predictions.

*B: An Analogy between Contour Maps and Equipotential Lines*

We designed materials to help students draw on ideas from topographical contour maps (Fig. 1) to predict the direction and magnitude of the electric field from electric equipotential lines (Fig. 2).  There were two important principles of electric potential we wanted to highlight with the common structure of the physics context and the analogy.

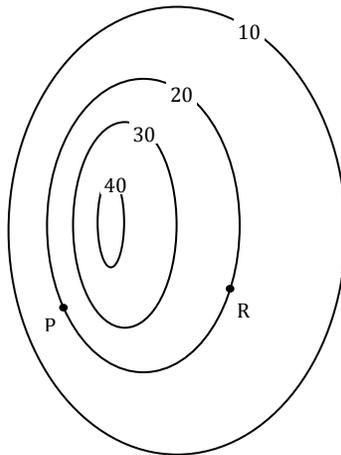

Figure 1: A topographical contour map representing an overhead view of a hill.  The lines represent locations of equal height, from 10 m to 40 m. (diagram adapted from the Open Source Tutorials in Physics Sensemaking)

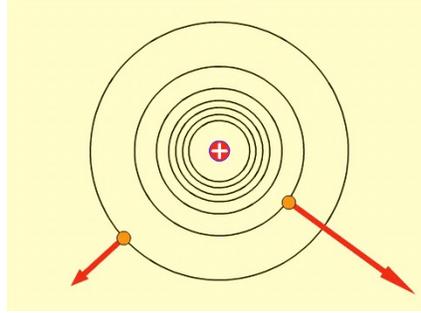

Figure 2: The equipotential lines for a positive point charge, drawn at intervals of every 5 V. The diagram shows the correspondence between equipotential lines and electric field by showing the direction and magnitude of the electric field at two points.

First, the direction of the electric field is perpendicular to the equipotential lines, pointing towards decreasing potential. In terms of the contour map, this is analogous to the direction a ball released on the hill will roll, perpendicular to the line of constant height, towards decreasing height. Second, the magnitude of the electric field is proportional to the density of the equipotential lines. For the contour map, the force pulling the ball down the hill is proportional to the steepness of the hill, indicated by density of the contour lines.

Previous research shows that students can have difficulties with these principles even after instruction. For example, it is common for students to incorrectly predict, even after instruction, that the magnitude of the electric field will depend on the value of the electric potential rather than how quickly the electric potential changes [34,35]. Again, we predict that *Case-by-Case* students will be drawn to attractive surface features like the value of the electric potential more often than students in the *General Rule* instruction.

*C: Materials*

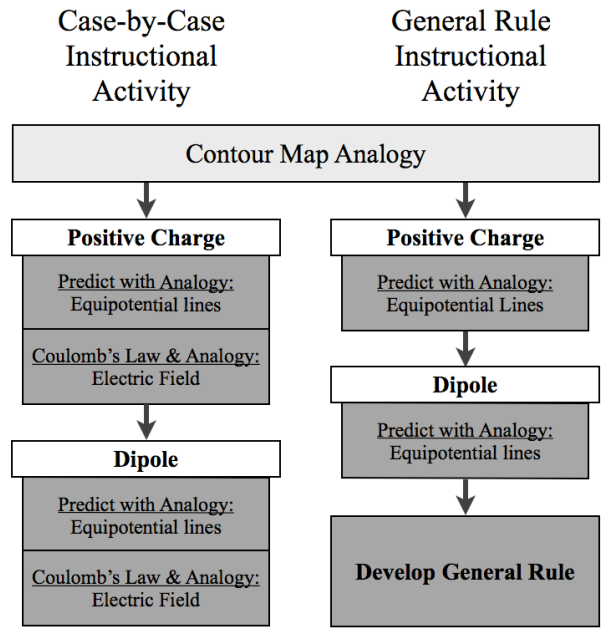

Figure 3: The sequence of activities for the *Case-by-Case* and *General Rule* instruction.

Our instructional goal was to have students connect an understanding of topographical contour maps to electric potential to predict the electric field from the equipotential lines, before they were taught about electric potential in class. The two instructional activity sequences contained in the worksheets that students received are shown in Figure 3. The full instructional materials are provided as supplementary materials.

Both sets of activities began with a brief introduction to contour maps – adapted from the Open Source Tutorial in Physics Sensemaking on electric potential [36]. Imagining a ball being placed on the hill, students were asked to draw the initial direction of motion and rate the relative steepness for two points on the hill, along with several questions that engaged their understanding of work done and change in potential energy along different paths on the hill. The worksheet then describes how the contour lines for hills are analogous to electric equipotential lines.

From here, the two instructional sequences diverged in how students were directed to map the analogy into our target physical context. In the next sub-sections, we describe the details of the two instructional sequences.

*i) Case-by-Case* instructional sequence:

The *Case-by-Case* students were led through a series of questions asking them to relate the contour map analogy to electric potential lines. The goal of this activity was for students to connect the contour map analogy to the target physical context by having students use the analogy to make predictions in electrostatics.

Positive Charge:
They were first asked to predict what the equipotential lines for a positive charge would be if drawn for 10, 20, 30, and 40 V. They were explicitly directed with the following hint: "Use the contour map analogy – imagine which way a positive charge should travel at different points and how steep the 'hill' must be."

They were then asked to use both (i) Coulomb's Law, which they had recently covered in the course, and (ii) the contour map analogy to identify the strength of the electric field everywhere along the 20 V line and compare the strengths of the electric field at the 20 V line to the electric field at the 10 V line.

The purpose of directing students to give the answer using both Coulomb's Law and the contour map analogy was to illustrate how the electric field could be determined in two, independent ways: by the charges or by the equipotential lines. The connection to Coulomb's Law provided students with additional scaffolding by connecting this new material to a topic they had recently learned, allowing students to check their contour map predictions against a more familiar Coulomb's Law prediction.

Following this, students were directed to use the Charges and Fields PhET simulation [37] to check their predictions. The simulation provided feedback about whether or not students' predictions were correct.

Dipole:
A similar sequence of questions for a dipole followed. Students predicted the shape of the equipotential lines with the contour map analogy. They were then again asked to use both (i) Coulomb's Law and (ii) the contour map analogy and the equipotential lines to predict the electric field strength and direction all along the 0 V line and the electric field direction at every point on the +5 V line, again using both Coulomb's law and the contour map analogy. Once again, they checked their predictions in the PhET simulation.

The overall sequence was designed to give students an easier situation (the positive charge) before a more complicated one (the dipole). The symmetry of the positive charge case and the analogy of the positive charge as "the top of a hill" make the shape of the equipotential lines easier to predict than in the case of the dipole. Similarly, the dipole was meant to serve as a contrast to the positive charge in terms of how the electric field behaves in relation to the equipotential lines. For the dipole, the direction of the electric field is not always pointing directly towards or away from one of the charges, as it is when considering the positive charge alone. Additionally, the magnitude of the electric field is not constant at all points along an equipotential line, as it is in the case of the positive charge. These different examples, in combination with the feedback provided by the PhET simulation, could help students see the important features of the equipotential lines needed for determining the electric field precisely. Yet, our prediction is that, since we are directing them to consider the questions in isolation and not all together, CC students would not experience the full benefit of these contrasts.

*ii) General Rule* (GR) instructional sequence:

Rather than leading students through questions on the direction and magnitude of the electric field for two different charge configurations, the *General Rule* instruction was asked students to come up with a general rule for how to determine the electric field from the equipotential lines.

Positive Charge and Dipole:
GR students were asked to use the analogy of contour maps to predict the shape of the equipotential lines for both the positive charge and the dipole, just as the CC students were. However, after making their predictions, a short paragraph made the distinction between checking to confirm your answer and testing your idea in lots of ways, explaining surprising results. It suggested that people learn the most by testing their ideas and explaining a surprising outcome. Students then used the Charges and Fields PhET simulation to draw the equipotential lines, checking their predictions, but they were also explicitly directed to explain something that was initially surprising in the simulation.

Develop General Rules:
Instead of making specific predictions with Coulomb's Law and the contour map analogy, students were then directed to use the PhET simulation to come up with general rules for answering the following two questions:

- How do you know the *direction of the electric field* from the equipotential lines? Explain. (Hint: Use the contour map analogy)

- How do you know the *strength of the electric field* from the equipotential lines? Explain. (Hint: Use the contour map analogy)

The worksheet provided no additional suggestions of what charge configurations to consider or what features to check. Once these rules were generated, students were told to use the remaining time to build different charge configurations in the simulation to test their two rules.

Both CC and GR instructional materials have the students map the analogy of contour maps over to equipotential lines, though in different ways. The key distinction between the two instructional sequences is that the GR activity pushes students to come up with general rules but does not provide explicit guidance on what specific instances to examine, whereas the CC students are directed to use the analogy to examine particular examples, but in isolation rather than all together. Additionally, the presence of the charges in the given cases allows use of Coulomb's law. Although this provides useful scaffolding in an unfamiliar task, CC students may over-rely on the location of the charges in determining the electric field, accomplishing the task without preparing themselves to make future predictions from only the equipotential lines.

*D: Assessment measures: pre-, mid-, and post-tests*

All assessment items require students to conceptually understand of how the direction and/or magnitude of the electric field can be read from the equipotential lines. The pre-test items come from four items on the Conceptual Survey in Electricity and Magnetism [34] covering the relation between equipotential lines and electric field. The mid- and post-test (shown in Figures 4 and 5, respectively) each contained two questions created by the researchers to evaluate what the instructional activities were designed to teach: how to determine the direction or magnitude of the electric field from the equipotential lines. Importantly, all assessment items display the equipotential lines without revealing the charge distribution generating those lines, so students cannot use Coulomb's Law to determine the electric field.

What is the direction of the electric field at point Z for the picture to the right?

A. ↖

B. ↗

C. ↘

D. ↙

E. None of the above.

Answer: _______

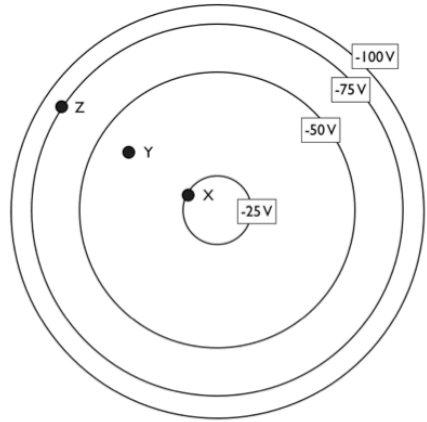

How does the strength of the electric field compare at points X, Y, and Z?

A. $E_X = E_Y = E_Z$
B. $E_X > E_Y > E_Z$
C. $E_X < E_Y < E_Z$
D. $(E_X = E_Y) > E_Z$
E. $(E_X = E_Y) < E_Z$
F. Not enough information

Answer: _______

Figure 4: Direction mid-test question and magnitude mid-test question asking students to use the equipotential lines to predict the electric field.

Which arrow is closest to the direction of the E-field at point A?

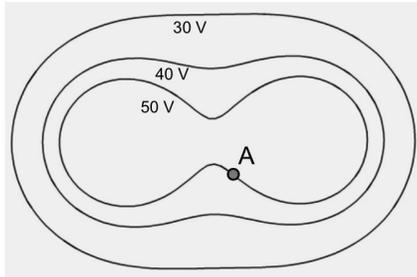

A) ↙
B) ↘
C) ↖
D) ↗
E) ↓

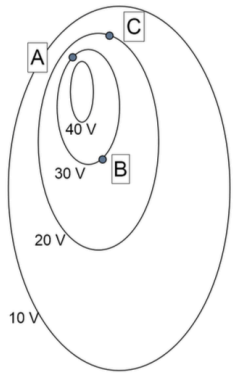

How does the magnitude of the electric field at these three points compare?

A. $(E_A = E_B) > E_C$
B. $E_C > (E_A = E_B)$
C. $E_A > E_B > E_C$
D. $E_A > E_C > E_B$
E. Not enough information

Figure 5: Direction post-test question and magnitude post-test question asking students to use the equipotential lines to predict the electric field.

Although there are differences between the pre-test items from the CSEM and the mid- and post-test items designed for this study, they all require an understanding of how electric potential is related to electric field. Though coarse, the CSEM pre-test helps us eliminate differences in prior physics knowledge as an explanation for any conditional effects.

*E: Research design sequence*

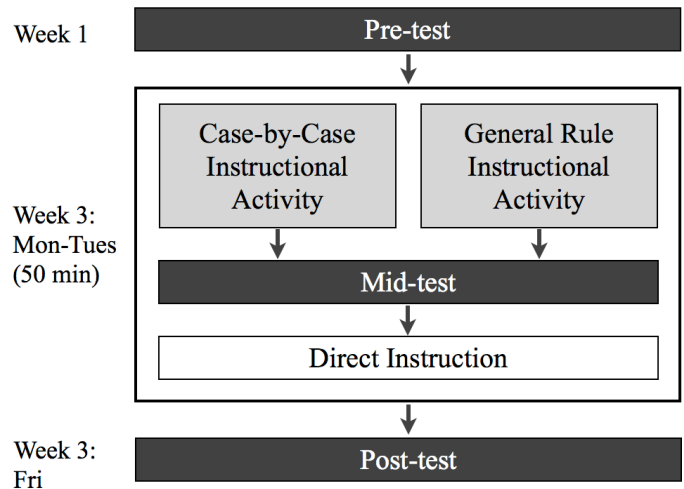

Figure 6: The research design sequence.

Figure 6 shows the sequence of the experimental design. During the first week of the course, all students were given the pre-test.

The main intervention occurred in discussion sections during the 3$^{rd}$ week of class, before students received direct instruction on the connection between electric potential and electric field. All discussion sections occurred on Monday and Tuesday, each lasting for 50 minutes. For 35 minutes, students either worked on the *General Rule* (GR) or *Case-by-Case* (CC) activities in groups of 3 or 4. Students were not allowed to look up the relevant principles in their textbook or online, because the instructional goal was for students to learn the physics principle from the analogy. Instead, they worked together as a group, using the simulation to answer the questions on the worksheet. In debriefing the TAs, they reported that both GR and CC students generally completed the activity, although the CC activity seemed a bit longer.

Then, students spent 5 minutes individually completing the mid-test. After the mid-test, students received direct instruction on the relation between equipotential lines and electric fields. Students were given a one-page summary to read on what they should have seen in the simulation, showing the equipotential lines and electric field at different points for the positive charge case and the dipole case. This summary also explicitly stated how the direction and magnitude of the electric field depend on the equipotential lines, illustrating these with the positive charge and dipole cases. TAs used the remaining time to either answer questions or present a lecture on the topic that they prepared on their own.

The post-test was embedded as clicker questions in the lecture on the Friday of the 3$^{rd}$ week of class. Before class, students were assigned textbook readings on the topic of electric potential. Leading up to the post-test, the lecturer asked indirectly related clicker questions and provided formal and mathematical explanations of electric potential and electric field. For the post-test questions, students were given about a minute to answer each clicker question individually. There was no instruction in the lecture that could be used to answer these post-test questions, so they serve as measures of the effect of the direct instruction in the discussion section and the assigned textbook readings. For the pretest, the discussion section activity, the mid-test, and the post-test, students were only awarded class credit for completion, not for correct answers.

To test our first hypothesis, that *GR better supports discovery and application of the correct principle than CC*, we compared the pre-test conditional differences and mid-test conditional differences, analyzing just the effect of the two instructional activities. To test our *competing predictions of the effect of later direct instruction*, we compared mid-test conditional differences to post-test conditional differences, measuring the effect of the direct instruction on the performance difference between the instructional groups.

**V*: Results and Analysis*

In our analysis, only students who completed the pre-, mid-, and post-test of our study were included (N = 231). This was primarily limited by the number of students who attended lecture, as only 57% of students that attended and completed a discussion activity completed the in-lecture post-test.

*A: Pre-test results*

On the pre-test questions, there was no difference in mean score (out of 4 possible points: $m_{GR} = 1.30$, $sd_{GR} = 1.08$, $m_{CC} = 1.45$, $sd_{CC} = 1.09$), $t(229) = 1.06$, $p = .29$. For each of the four questions, there was no significant difference in correctness by condition (all $p > .10$). These results show no significant difference in prior knowledge before the course, with a slightly higher score for the CC students.

*B: Mid-test results*

Because there was no significant difference between the two conditions at pre-test, we use just the mid-test results to illustrate the effect of the instructional activities. Figure 7 shows the percentages of students who correctly answered the direction and magnitude mid-test questions. Examining the first hypothesis with the mid-test direction and magnitude scores, there is a 24% difference in correctness by condition for magnitude, $\chi^2(1, N=231) = 12.7$, $p < .001$, but no difference for direction, $\chi^2(1, N=231) < .1$.

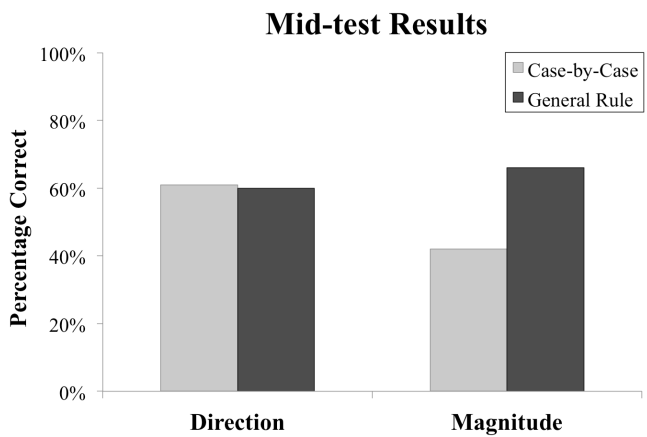

Figure 7: Correctness on the direction and magnitude mid-test questions. Between the CC and GR conditions, there is no difference on the direction question, but an advantage of GR on the magnitude question.

Why does this advantage for GR appear on the magnitude question and not the direction question? We argue that for the particular questions asked, incorrect, though plausible, case-specific reasoning can lead to a correct answer on the direction question but not the magnitude question. Therefore, the instructional difference between GR and CC materials would be more apparent on the magnitude question.

To illustrate this, we show the distribution of student responses on the magnitude mid-test question, shown in Table 1. On the magnitude question, the correct reasoning is that the electric field magnitude depends on the density of equipotential lines. Therefore, the correct answer is (C), $E_X < E_Y < E_Z$. In this case, the most common incorrect answer in both conditions is (B), $E_X > E_Y > E_Z$. CC students give the common incorrect answer twice as often as GR students.

| | Magnitude mid-test question | | |
|---|---|---|---|
| | Correct: Proportional to density of equipotential lines ($E_X < E_Y < E_Z$) | Common Incorrect ($E_X > E_Y > E_Z$) | Other responses |
| **General Rule** | 66% | 28% | 6% |
| **Case-by-Case** | 42% | 55% | 3% |

Table 1: Student response percentages to the mid-test questions

CC students may come to this incorrect answer by matching the mid-test questions to the positive charge case by virtue of the common concentric circle geometry of the equipotential lines. For example, from the positive charge case alone, a student could reasonably, though incorrectly, conclude that the relative strength of the electric field at different points always covaries with either the value of the electric potential or the distance from the center of the equipotential lines. These surface features may be more visually salient to students than the important second-order feature of how densely packed the equipotential lines are. Applying either of these incorrect conclusions to the magnitude mid-test question leads to the common incorrect answer, $E_X > E_Y > E_Z$. By explicitly articulating a general rule, GR students are more likely to attend to the key features of the analogy, comparatively minimizing use of these common surface-feature-based principles.

However, for the direction mid-test question, similar incorrect, case-specific reasoning may not be penalized. On the direction mid-test question, the correct answer (A) is that the electric field points radially away from the center of the concentric equipotential lines, because the electric field points perpendicularly to the equipotential line, towards decreasing potential. Yet, students who draw similar incorrect conclusions from the positive charge case would also select the correct answer here, by coincidence. For example, a surface-feature-based explanation that the electric field always points away from the center of circular equipotential lines matches both the positive charge case and the correct answer to the direction mid-test question. By this argument, the results from the direction mid-test question are likely an overestimate of how many students were using the correct rule, potentially masking differences between GR and CC. This is not the case for the direction post-test question, as it has a different equipotential line geometry than either the positive charge or dipole case investigated in the CC instruction.

*C: Post-test results*

In order to investigate the competing hypotheses of the effect of subsequent direct instruction after the activity, we compare the mid-test condition differences to the post-test condition differences. The percentage of correct responses on the direction and magnitude post-test questions are shown in Figure 8. We compare the post-test differences to the 24% difference on the magnitude mid-test question, which we have argued represents the difference between the two instruction activities before direct instruction. The difference between GR and CC performance on the mid-test and post-test questions is shown in Table 2. As noted previously, the direction mid-test results are anomalous and cannot be directly compared to other differences.

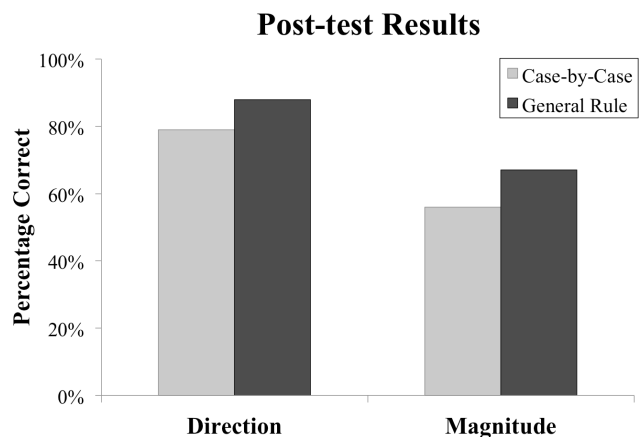

Figure 8: Correctness on the direction and magnitude post-test questions. On both the direction and magnitude questions, GR performs 10% better than CC.

|  | **Difference in % Correct: (GR – CC)** | |
| --- | --- | --- |
|  | Direction | Magnitude |
| Mid-test | -1% | 24% |
| Post-test | 9% | 11% |

Table 2: The difference by condition (GR – CC) of percentage of correct responses on the mid-test and post-test questions.

As established in the previous section, the magnitude mid-test question shows a significant difference between GR and CC instruction for success in applying a rule for determining the E-field magnitude from the equipotential map. Using the Mantel-Haenszel test to consider the direction and magnitude post-test questions together, there is still a significant difference at post-test between conditions, with GR students outperforming CC students, $\chi^2_{MH}(1, N = 462) = 5.35, p = .021$. For the direction question, GR outperforms CC by 9% at post-test, which is marginally significant, $\chi^2(1, N=231) = 3.39, p = .065$. For the magnitude questions, the 11% difference between GR and CC at post-test is also marginally significant, $\chi^2(1, N = 231) = 2.70, p = .10$.

The maintained difference at post-test between GR and CC is not due to a lack of improvement by CC students. CC students improved their performance from mid- to post-test by 18% for the direction questions, McNemar's test: $p = .002$, and by 14% for the magnitude questions, McNemar's test: $p = .033$. This shows that the additional benefit of the GR activity, which we argue helps students see the deep structure beneath the case-specific surface features, remains even after direct instruction increases performance.

## VI: Summary

Students who sought a general rule in the instructional activity were more successful at discovering and applying the relationship between electric potential and electric field lines. Analysis of the specific items lends some insight to the possible mechanism: students led to consider individual cases in isolation were vulnerable to making surface feature-based predictions. We argue that the GR students outperformed CC students after the instructional activity because they became better at attending to the features relevant to the correct physics principle.

Overall, there is a significant advantage for GR on the post-test questions, even though there is only a marginal difference on either the direction or magnitude question alone. Interestingly, both of our competing hypotheses here were incorrect. The difference between conditions was neither nullified nor enhanced. Instead, the conditional difference persisted beyond direct instruction but was diminished.

More broadly, this study illustrates that a common instructional maxim, "the best way to help students understand a new idea is to provide scaffolded practice," should be interpreted with care. We showed that a step-by-step guided series of questions, meant to coherently illustrate to students the connections between electric fields and electric potential lines, was less helpful than having students generate rules with little guidance as to which specific cases to consider. Attempting to show students what is important through a series of specific cases may actually end up limiting their perspective.

Overall, these results provide some evidence for the instructional efficacy of having students develop general rules in physics from analogous domains. However, this study alone does not determine general best design practices for physics instructional materials. These benefits may depend on the specific measures of success. The questions on the mid-test asked students to infer the electric field from the equipotential lines. However, this understanding of electric field and electric potential alone does not represent a full conceptual and quantitative understanding of these topics. The *Case-by-Case* analysis could be beneficial for developing a more robust connection between Coulomb's law and electric potential, for example. The advantage of the *General Rule* instruction may also depend on the nature of the particular target physics principles developed through analogy. For some concepts, practice in applying the rules to different situations may be more important than the statement of a general rule. More work is needed to understand how the benefit of seeking a general rule through analogy is conditional on the physics content to be learned and the kinds of questions used to assess that learning.

Our result that the development of a general rule does not help students gain more from direct instruction than the CC activity is surprising in light of the research showing that these kinds of generalization activities can prepare students for future learning from such direct instruction. One reason for this surprising result could be that our mid-test and post-test questions measured relatively near transfer of the same physics concepts on similar problems, whereas other studies investigated further transfer. Studies showing that these activities can prepare students for future learning from direct instruction tend to look at student understanding beyond the original generated rule and beyond the apparent content of the direct instruction. It could be that there is still some unmeasured benefit of activities like GR on future learning that would be evident on different kinds of learning tasks.

**VII: Conclusion**

Beyond the specific outcomes of our study, our goal here is to illustrate the consequentiality of instructional details not specified by broader research-based instructional design principles such as "interactive engagement". Research focused on the learning benefits of interactive engagement does not necessarily suggest how to design effective instructional materials. As the types of instructional design differences at the level of *General Rule* vs. *Case-by-Case* are not often explicitly addressed, instructors have little guidance on detailed instructional design decisions and the resulting instruction may be far from optimum. Leaving such details unattended may also unfortunately suggest to instructors and curriculum designers that these kinds of instructional details are unimportant.

Understanding the impact of these different instructional designs has the potential to feed into many aspects of physics instruction. Knowledge of how the instructional details affect student learning could guide consistent design of instructional materials from a set of basic principles, decreasing reliance on any one curriculum designer's instructional wisdom.  A better understanding of these instructional details can also illuminate the critical features of existing instructional materials.  For example, existing effective materials that incorporate analogical instruction may share some key structural features, such as directing students to explicitly map a general rule from an analogical domain to the target physics domain. One direction for future research is the augmentation of studies that investigate the general efficacy of PER-inspired materials with studies of how slight modifications to the structure of those materials are consequential for student reasoning and learning outcomes, revealing what structural features contribute to the instructional success of good materials.

One challenge in adopting PER-based instructional methods is that instructors often want to modify or adapt instructional materials to fit local classroom contexts. However, without guidance as to what components of the materials are flexible and what components are critical, instructors may make changes that subvert the efficacy of these materials.  Knowledge of the consequentiality of the instructional details may not only lead to design of effective instructional materials, but also effective adaptation of those materials to different instructional contexts, possibly supporting instructor success with and buy-in of novel PER-based instructional methods.

**Acknowledgements:**

We would like to thank members of the AAALab for insightful comments on the design of the instructional materials. This work was supported by the Gordon and Betty Moore Foundation.